\documentclass[%
twocolumn,
preprintnumbers,
superscriptaddress,
nofootinbib,
reprint,
showkeys,
showpacs,
amsmath,amssymb,
aps,
]{revtex4-1}

\usepackage{graphicx}
\usepackage{dcolumn}
\usepackage{bm}
\usepackage{amsfonts,amsbsy}
\usepackage{aas_macros}
\usepackage{mathrsfs}
\usepackage{url}
\usepackage{multirow}
\usepackage{xcolor}
\usepackage{color}
\usepackage{ulem}
\usepackage{enumerate}
\usepackage{textcmds}
\usepackage{mathtools}
\usepackage{booktabs}
\usepackage{tabularx}
\usepackage[linkcolor=blue,citecolor=blue,backref=page]{hyperref} 
\usepackage{epstopdf}

\renewcommand{\vec}[1]{\boldsymbol{#1}}

\newcolumntype{p}{D{,}{\pm}{-1}}

\begin{document}

\preprint{Version \today}

\title{Ultrahigh-energy diffuse gamma-ray emission from cosmic-ray interactions with medium surrounding acceleration sources}

\author{Pei-pei Zhang}
 \affiliation{Key Laboratory of Dark Matter and Space Astronomy, Purple Mountain Observatory, 
Chinese Academy of Sciences, Nanjing 210023, China}
\affiliation{School of Astronomy and Space Science, University of Science and Technology of China, 
Hefei 230026, Anhui, China}
\affiliation{Hebei Normal University, Shijiazhuang 050024 , Hebei, China}
\affiliation{Key Laboratory of Particle Astrophysics, Institute of High Energy Physics, 
Chinese Academy of Sciences, Beijing 100049, China}

\author{Bing-qiang Qiao}\email{qiaobq@ihep.ac.cn}
\affiliation{Key Laboratory of Particle Astrophysics, Institute of High Energy Physics, 
Chinese Academy of Sciences, Beijing 100049, China}

\author{Qiang Yuan}
\affiliation{Key Laboratory of Dark Matter and Space Astronomy, Purple Mountain Observatory, 
Chinese Academy of Sciences, Nanjing 210023, China}
\affiliation{School of Astronomy and Space Science, University of Science and Technology of China, 
Hefei 230026, Anhui, China}

\author{Shu-wang Cui}
\affiliation{Hebei Normal University, Shijiazhuang 050024 , Hebei, China}

\author{Yi-qing Guo}\email{guoyq@ihep.ac.cn}
\affiliation{Key Laboratory of Particle Astrophysics, Institute of High Energy Physics, 
Chinese Academy of Sciences, Beijing 100049, China}
\affiliation{University of Chinese Academy of Sciences, Beijing 100049, China}

\date{\today}

\begin{abstract}
The diffuse $\gamma$-ray spectrum at sub-PeV energy region has been measured
for the first time by the Tibet-AS$\gamma$ experiment. It will shed new light
on the understanding of the origin and propagation of Galactic cosmic rays at
very high energies. It has been pointed out that the traditional cosmic ray
propagation model based on low energy measurements undershoot the new data, 
and modifications of the model with new ingredients or alternative propagation 
framework is required. In this work, we propose that the hadronic interactions 
between freshly accelerated cosmic rays and the medium surrounding the sources, 
which was neglected in the traditional model, can naturally account for the 
Tibet-AS$\gamma$ diffuse emission. We show that this scenario gives a consistent 
description of other secondary species such as the positron spectrum, the 
boron-to-carbon ratio, and the antiproton-to-proton ratio. As a result, the 
electron spectrum above 10 TeV will have a hardening due to this secondary 
component, which may be tested by future measurements.
\end{abstract}

\pacs{Valid PACS appear here}

\maketitle

\section{Introduction}
\label{sec:intro}
The Galactic diffuse $\gamma$-ray emission (DGE) is expected to be
produced by interactions between cosmic rays (CRs) and the interstellar
medium (ISM) as well as the interstellar radiation field (ISRF), during
the propagation of CRs in the Milky Way. The DGE includes mainly three
components: the decay of $\pi^0$ from inelastic hadronic interactions
between CR nuclei and the ISM, the bremsstrahlung of CR electrons and
positrons (CREs) in the ISM, and the inverse Compton scattering (ICS)
component of CREs scattering off the ISRF \citep{2007ARNPS..57..285S}.
This model can consistently describe most of the DGE data below 100 GeV
and the locally observed results of CRs \citep{2004ApJ...613..962S,
2010ApJ...720....9Z,2012ApJ...750....3A}, with only slight excesses in
the inner Galactic plane which was suggested to be due to unresolved
sources or spectral variations of CRs throughout the Milky Way
\citep{2012ApJ...750....3A}. Measurements of DGE at higher energies
are thus very important to further test the model.

The ground based experiments Milagro and ARGO-YBJ measured the DGE above
TeV energies, for a few selected sky regions along the Galactic plane
\citep{2007ApJ...658L..33A,2008ApJ...688.1078A,2015ApJ...806...20B}.
Particularly, in the Cygnus region, the Milagro observation identified
an excess \citep{2007ApJ...658L..33A} compared with the CR propagation
model tuned to account for the low-energy DGE \citep{2004ApJ...613..962S}.
Some fresh sources in such a region may explain the excess
\citep{2009ApJ...695..883B,2016ChPhC..40k5001G}. Very recently, the DGE
in the Galactic plane above 100 TeV energies was for the first time
measured by the Tibet-AS$\gamma$ experiment \citep{2021PhRvL.126n1101A},
which has attracted wide attention for possible physical discussion
\citep{2021arXiv210315029K,2021arXiv210402838D,2021arXiv210409491F,
2021ApJ...914L...7L,2021arXiv210403729Q,2021PhyOJ..14...41H,
2021arXiv210501826E,2021arXiv210500959K,2021arXiv210513378B,
2021arXiv210507242D,2021arXiv210506151L,2021Univ....7..141T,
2021arXiv210505680N}. 
The Tibet-AS$\gamma$ fluxes are higher than the prediction of the
conventional CR propagation model, and additional components or
modification of the conventional propagation framework may be needed
\citep{2021PhRvL.126n1101A,2021ApJ...914L...7L,2021arXiv210403729Q}.

However, in the traditional DGE modeling the secondary particle production
(including $\gamma$ rays) due to interactions between newly accelerated CRs 
and the gas surrounding the sources is usually omitted. This component may
not be small \citep{2019PhRvD.100f3020Y}, and is expected to be more and more
important at high energies since the CR spectra around the sources are harder
than those diffusing out in the Milky Way. The possible confinement of CRs
around the sources may further enhance this component of secondary particles.
In this work, we investigate this scenario in light of the ultrahigh-energy
(UHE) diffuse emission measured by Tibet-AS$\gamma$. The consequence of
such interactions for other types of secondary particles, such as the
B/C ratio, the positron and antiproton fluxes will also be investigated.
We confront this model with the up-to-date measurements of $\gamma$ rays
and CRs, and find a consistent description of these new data.

\section{Model Description}
\label{sec:model}
\subsection{Propagation of CRs}
It has been recognized in recent years that the propagation of CRs in the
Milky Way should depend on the spatial locations, as inferred by the HAWC
and LHAASO observations of extended $\gamma$-ray halos around pulsars
\citep{2017Sci...358..911A,2021PhRvL.126x1103A} and the spatial variations 
of the CR intensities and spectral indices from Fermi-LAT observations 
\citep{2016PhRvD..93l3007Y,2016ApJS..223...26A}. The spatially-dependent 
propagation (SDP) model was also proposed to explain the observed hardenings 
of CRs \citep{2012ApJ...752L..13T,2015PhRvD..92h1301T,2016PhRvD..94l3007F,
2016ApJ...819...54G,2018ApJ...869..176L,2018PhRvD..97f3008G,
2020ChPhC..44h5102T}, and also the large-scale anisotropies with the
help of a nearby source \citep{2019JCAP...10..010L,2019JCAP...12..007Q,
2020FrPhy..1624501Y}.

In the SDP model, the diffusive halo is divided into two parts, the
inner halo (disk) and the outer halo. In the inner halo, the diffusion
coefficient is much smaller than that in the outer halo, as indicated
by the pulsar halo observations. 
{To enable a smooth variation of the diffusion coefficient 
$D_{xx}$, we parametrize it as}
\begin{equation}
D_{xx}(r,z, {\cal R} )= D_{0}F(r,z)\beta^{\eta} \left(\dfrac{\cal R}
{{\cal R}_{0}} \right)^{\delta_0 F(r,z)},
\label{eq:diffusion}
\end{equation}
where $r$ and $z$ are cylindrical coordinate, ${\cal R}$ is the particle's
rigidity, $\beta$ is the particle's velocity in unit of light speed,
$D_0$ and $\delta_0$ are constants representing the diffusion coefficient
and its high-energy rigidity dependence in the outer halo, $\eta$ is a 
phenomenological constant in order to fit the low-energy data.
{The spatial dependence function $F(r,z)$ is given as
\citep{2020ChPhC..44h5102T},
\begin{equation}
F(r,z) = \left\{
\begin{array}{ll}
{g(r,z)+[1-g(r,z)]} \left(\dfrac{{z}}{\xi{z}_{\rm h}} \right)^{n},  &  {{|z|} \leq \xi{z}_{\rm h}} \\
\\
1,  &  { {|z|} > \xi{z}_{\rm h}} \\
\end{array},
\right.
\end{equation}
in which $g(r,z)$=$N_m$/[1+$f(r,z)$], and $f(r,z)$ is the source density 
distribution (see below Sec. II. B), $z_h$ is the half-thickness of the 
propagation cylinder, and $\xi z_{h}$ is the half-thickness of the inner halo.
The factor $\left(\frac{{z}}{\xi{z}_{\rm h}} \right)^{n}$ describes the smoothness 
of the parameters at the transition between the two halos. Note that the spatial
dependence of the diffusion coefficient is phenomenologically assumed. Physically
it may be related with the magnetic field distribution, or possibly the turbulence 
driven by CRs \citep{2012PhRvL.109f1101B}. The model parameters used in this 
work are listed in Table I.}
We adopt the diffusion reacceleration model in this work, with the reacceleration 
being described by a diffusion in the momentum space. The momentum diffusion
coefficient, $D_{pp}$, correlates with $D_{xx}$ via $D_{pp}D_{xx} = \dfrac{4p^{2}v_{A}^{2}}{3\delta(4-\delta^{2})
(4-\delta)}$, where $v_A$ is the Alfv\'en velocity, $p$ is the momentum,
and $\delta$ is the rigidity dependence slope of the spatial diffusion
coefficient \citep{1994ApJ...431..705S}. The numerical package DRAGON is
used to solve the propagation equation of CRs \citep{2017JCAP...02..015E}.
For energies smaller than tens of GeV, the fluxes of CRs are suppressed
by the solar modulation effect. We use the force-field approximation
\citep{1968ApJ...154.1011G} to account for the solar modulation.

\begin{table}[!htb]
\begin{center}
\begin{tabular}{cccccccccc}
\toprule[1.5pt]
 $D_0$  & $\delta$ & ${N}_m$ & $\xi$ & $n$ & $\eta$ & ${\cal R}_0$  & $v_A$  & $z_h$  \\
$[10^{28}$cm$^2$/s] & $ $ & $ $ & $ $ & $ $ & $ $ & [GV] & [km/s] & [kpc] \\
\hline
4.9  & ~0.55~  & ~0.57~ & ~0.1~ & ~4~ & ~0.05~ & 2 & 6 & 5\\
\bottomrule[1.5pt]
\end{tabular}
\end{center}
{
\caption{Propagation parameters of the SDP model.}
}
\end{table}

\subsection{Background source distribution} \label{subsec:bkg}
Supernova remnants (SNRs) are considered to be the most plausible candidates 
for the acceleration of CRs. The spatial distribution of SNRs are approximated 
as an axisymmetric form parametrized as
\begin{equation}
f(r, z) = \left(\dfrac{r}{r_\odot} \right)^\alpha \exp \left[-\dfrac{\beta(r-r_\odot)}{r_\odot} \right] \exp \left(-\dfrac{|z|}{z_s} \right) ~,
\label{eq:radial_dis}
\end{equation}
where $r_\odot \equiv 8.5$ kpc represents the distance from the Galactic
center to the solar system. Parameters $\alpha$ and $\beta$ are taken to be
$1.69$ and $3.33$ \citep{1996A&AS..120C.437C}. The density of the SNR 
distribution decreases exponentially along the vertical height
from the Galactic plane, with $z_{s} = 200$ pc.

{The injection spectrum of nuclei and primary electrons
are assumed to be an exponentially cutoff broken power-law function of 
particle rigidity ${\cal R}$
\begin{equation}\label{eq:spectrum_CRE}
q({\cal R}) = q_{0} \left\{
\begin{array}{lll}
 \left(\dfrac{{\cal R}}{{\cal R}_{\rm br}} \right)^{\nu_{1}},  &  {{\cal R} \leq {\cal R}_{\rm br}} \\
\\
\left(\dfrac{{\cal R}}{{\cal R}_{\rm br}} \right)^{\nu_{2}} \exp\left[-\dfrac{\cal R}{{\cal R}_{\rm c}} \right],  &  { {\cal R} > {\cal R}_{\rm br}} \\
\end{array}
\right.,
\end{equation}
where $q_{0}$ is the normalization factor, $\nu_{1,2}$ are the spectral
incides, ${\cal R}_{\rm br}$ is break rigidity, ${\cal R}_{\rm c}$
is the cutoff rigidity. The spectral break is employed to fit the low-energy
spectra of CRs, which is not the focus of the current work.}

\subsection{Local pulsar and local SNR}
\label{subsec:nySNR}
At TeV energies, CREs originate from sources within $\sim1$ kpc around
the solar system \citep{2018SCPMA..61j1002Y}. In this small region, the
hypothesis of continuous distribution may not be valid any more.
Studies show that the discrete effect of nearby CR sources could induce
large fluctuations, especially at high energies \citep{2011JCAP...02..031M,
2012A&A...544A..92B,2017ApJ...836..172F}. The contribution of nearby
sources to CREs has been studied in the past works
(see e.g., \citep{2012APh....39....2S, 2014JCAP...04..006D,
2017PhRvD..96b3006L,2018ApJ...854...57F}). In this work, we assume a
nearby pulsar to account for the positron excess above $\sim 20$ GeV.
The propagation of CREs injected instantaneously from a point source
is described by a time-dependent propagation equation
\citep{1995PhRvD..52.3265A}. The injection rate as a function of time
and rigidity is assumed to be
\begin{equation}
Q^{\rm psr}({\cal R},t)=Q^{\rm psr}_{0}(t) \left(\dfrac{\cal R}{{\cal R}_0}
\right)^{-\gamma} \exp \left[-\dfrac{\cal R}{{\cal R}^{\rm e_{\pm}}_{\rm c}}
\right] ~,
\label{eq:nearby}
\end{equation}
where ${\cal R}^{\rm e_{\pm}}_{\rm c}$ is the cutoff rigidity of its
accelerated CREs. A continuous injection process of electron and
positron pairs with injection rate proportional to the spindown
power of the pulsar is assumed, i.e.,
\begin{equation}
Q_0^{\rm psr} (t) \propto \dfrac{q_0^{\rm psr}}{\tau_0(1+t/\tau_0)^2 } ~,
\end{equation}
where $\tau_0$ is a characteristic time scale of the decay of the
spindown \citep{2010ApJ...710..958K, 2013PhRvD..88b3001Y}.

The progenitor of this pulsar produces an SNR, which may accelerate
primary nuclei and electrons during its early evolution stage.
This local source contribution of primary electrons may be necessary,
given the different spectral behaviors of positrons and electrons
\citep{2021JCAP...05..012Z}. The injection process of the SNR is
approximated as burstlike. The source injection rate is assumed to
be the same as Eq.~(\ref{eq:nearby}) but with
\begin{equation}
Q_{0}^{\rm snr}(t) = q_{0}^{\rm snr} \delta(t-t_0) ~, \\
\end{equation}
where $t_0$ is the time of the supernova explosion. The propagated
spectrum from the local pulsar and SNR is thus a convolution of the
Green's function and the time-dependent injection rate $Q_0(t)$
\citep{1995PhRvD..52.3265A}
\begin{equation}
\varphi(\vec{r}, {\cal R}, t) = \int_{t_i}^{t} G(\vec{r}-\vec{r}^\prime, t-t^\prime, {\cal R}) Q_0(t^\prime) d t^\prime .
\end{equation}
The normalization is determined through fitting Galactic cosmic rays 
energy spectra, which results in a total energy of $\sim 2.3\times 10^{50}$ 
erg for protons and $\sim 1.4\times 10^{50}$ erg for helium. If 10\% of 
kinetic energy is used to accelerate CRs, the total energy of supernova 
explosion is estimated to $\sim 3.7\times 10^{51}$ erg.
{Note that the local source introduced here is to account 
for the energy spectra of CRs, and is independent of the excess of the
UHE diffuse $\gamma$ rays.}

\subsection{Secondary particles from interactions of freshly accelerated CRs}

The freshly accelerated CRs at sources could also interact with the
surrounding gas before they escape from the source regions and enter the
diffusive halo. Secondary electrons, positrons, antiprotons, and $\gamma$
rays could be produced, whose yields can be calculated as
\begin{equation}
\begin{split}
Q_{{\rm sec},j}  \; = \;
\sum_{i = \rm p, He} {\displaystyle \int\limits_{E_{\rm th}}^{+ \infty}} \; d E_i \;
v \; \left\lbrace n_{\rm H}
{\displaystyle \frac{d \sigma_{i + {\rm H} \to j}}{d E_j }} \right. \\
\left. +n_{\rm He} {\displaystyle \frac{d \sigma_{i + {\rm He} \to j}}{d E_j }} \right\rbrace
Q_i(E_i) ~,
\label{sec_source}
\end{split}
\end{equation}
where $n_{\rm H,He}$ is the number density of hydrogen and helium,
$d\sigma_{i+{\rm H} \to j}/dE_j$ is the differential cross section of
the production of secondary particle $j$ from primary particle $i$.
The yields of secondary nuclei (such as boron) are simply
\begin{eqnarray}
Q_{{\rm B},j} = \sum_{i = \rm C, N, O} (n_{\rm H} \sigma_{i+{\rm H}\rightarrow j} +n_{\rm He} \sigma_{i+{\rm He}\rightarrow j} )v Q_i(E).
\end{eqnarray}
Secondary charge particles also propagate in the Galaxy, which are also
calculated with the DRAGON package.

\section{Results}
\label{sec:res}
\subsection{Spectra of CR nuclei}
\begin{figure*}[!htb]
\centering
\includegraphics[width=0.52\textwidth]{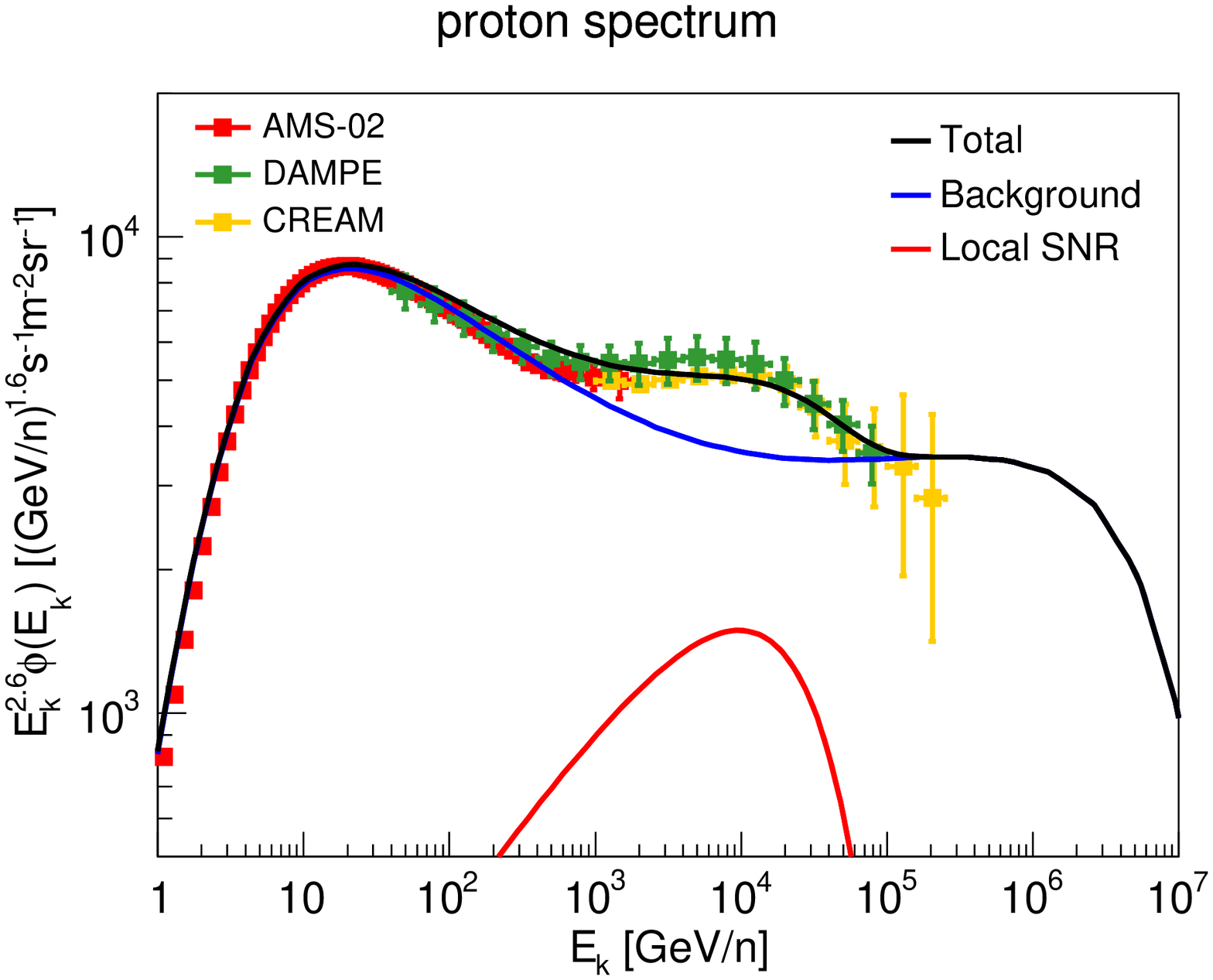}
\includegraphics[width=0.45\textwidth]{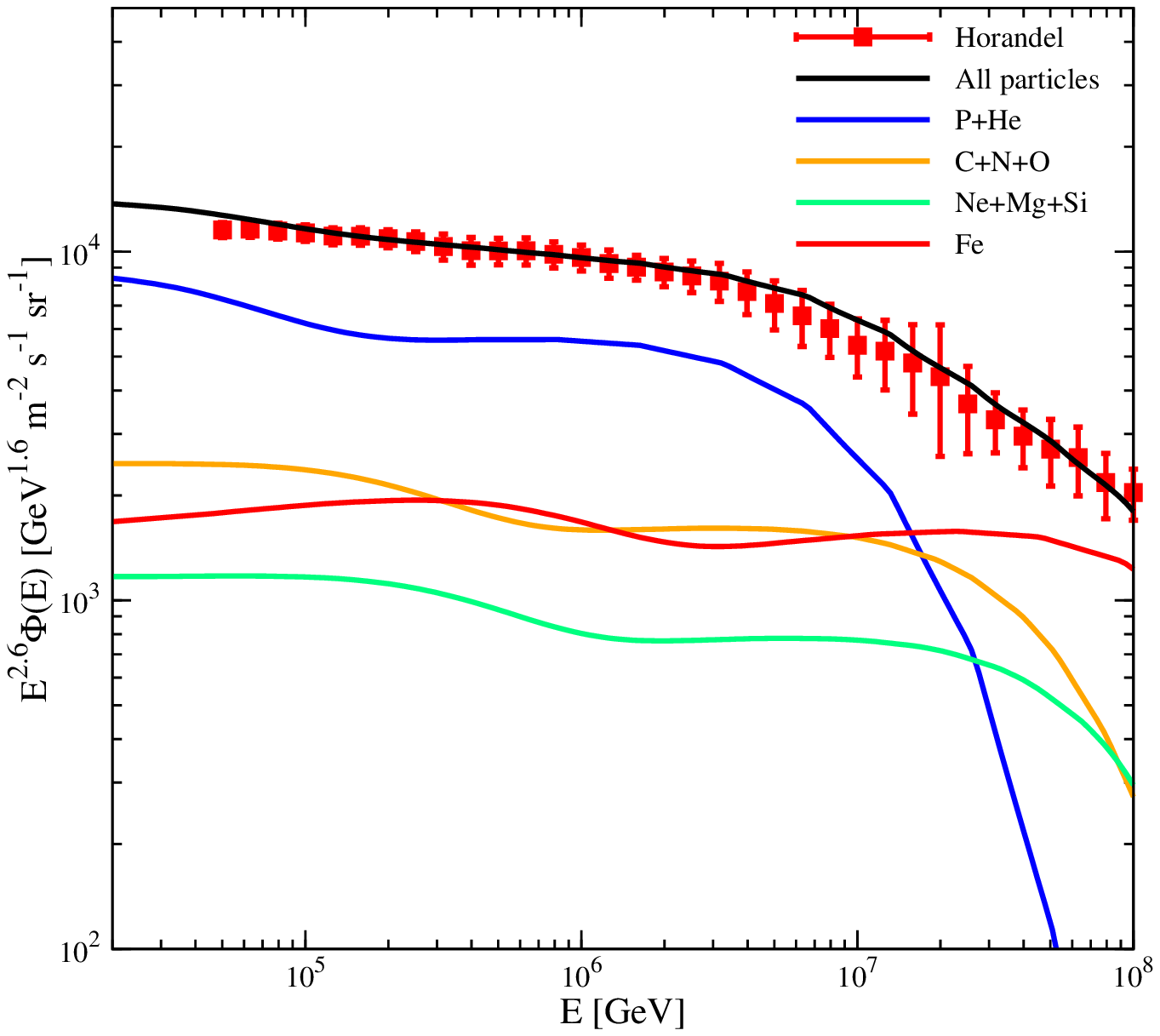}
\caption{The spectra of protons (left) and all particles (right).
The measurements of proton spectra are from AMS-02 \citep{2015PhRvL.114q1103A},
CREAM \citep{2017ApJ...839....5Y}, and DAMPE \citep{2019SciA....5.3793A}.
The all-particle spectrum is taken from the normalized result of
\citep{2003APh....19..193H}.}
\label{fig:crs-spect}
\end{figure*}

The left panel of Fig. \ref{fig:crs-spect} shows the proton spectrum
expected from the model, compared with the measurements
\citep{2015PhRvL.114q1103A,2017ApJ...839....5Y,2019SciA....5.3793A,
2003APh....19..193H}. The model parameters for different source
components are given in Table \ref{table-parm}. The hardening of the
proton spectrum around several hundred GeV can be attributed to the
summation of the background contribution and the local SNR contribution,
and the softening around 14 TeV is mainly due to the spectral cutoff
of the local SNR. Similar spectral features are expected to be present
for all species, as revealed recently by the DAMPE helium spectral
measurement \citep{2021PhRvL.126t1102A}. In the right panel of
Fig. \ref{fig:crs-spect} we show the total spectrum of high-abundance
nuclei, compared with the data \citep{2003APh....19..193H}. For the
parameters we adopt, the knee of the all-particle spectrum is mainly
due to the spectral cutoff of protons and helium nuclei from the
background SNRs.

\subsection{Diffuse $\gamma$ rays}

\begin{figure*}[!htb]
\centering
\includegraphics[width=0.48\textwidth]{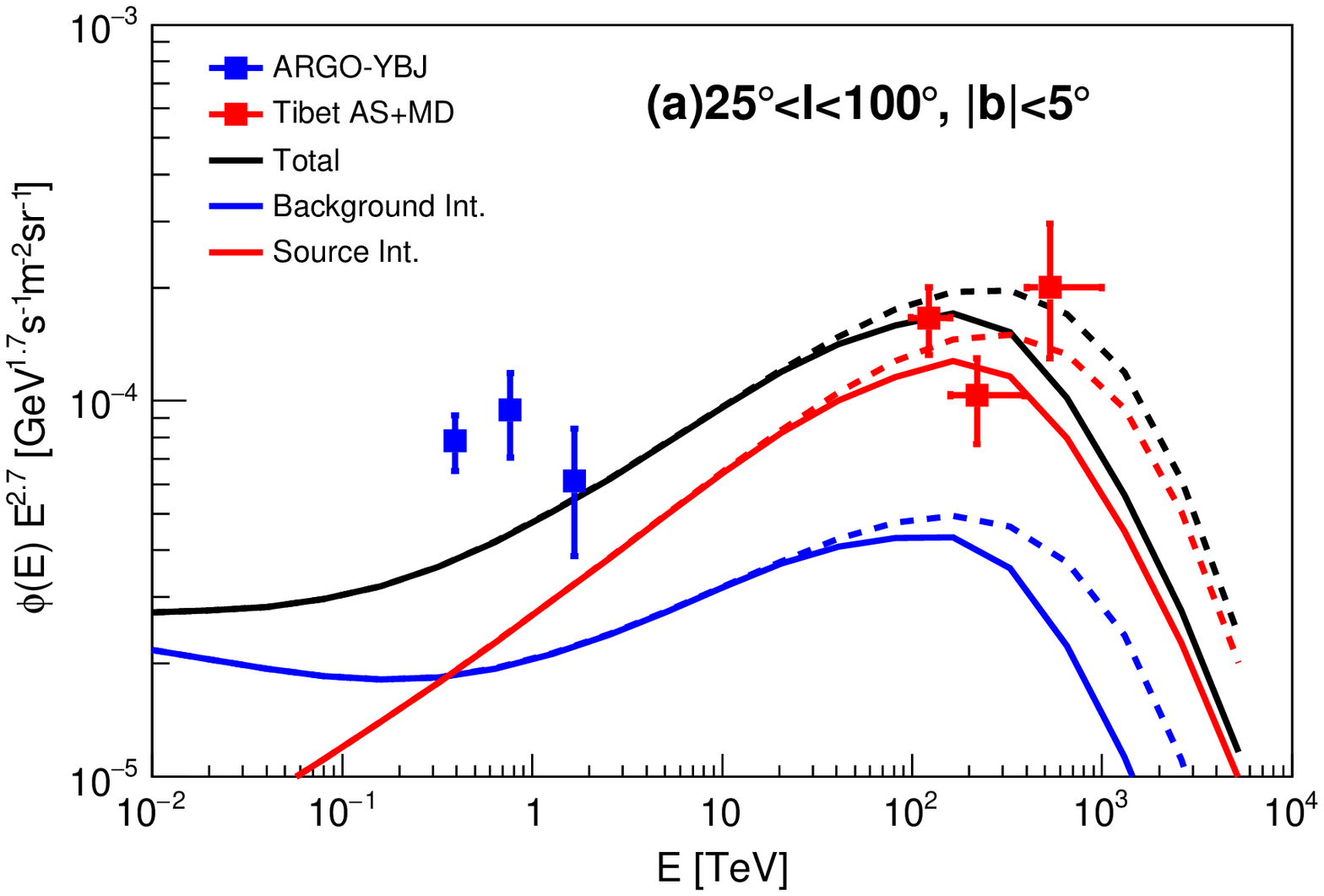}
\includegraphics[width=0.48\textwidth]{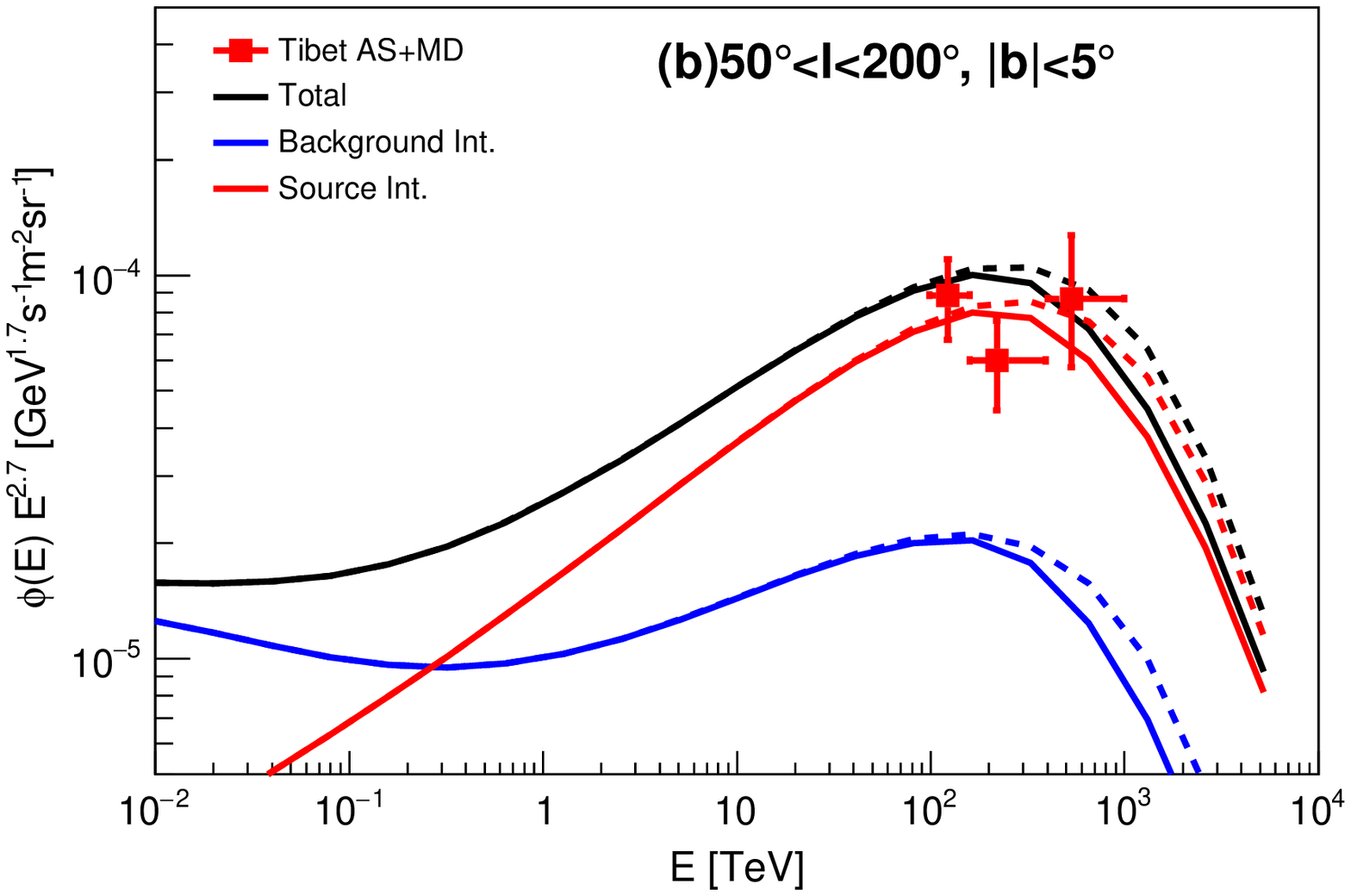}
\caption{Diffuse $\gamma$-ray spectra from the model calculation, compared
with the measurements by ARGO-YBJ \citep{2015ApJ...806...20B} and Tibet
AS$\gamma$ \citep{2021PhRvL.126n1101A}. {The dashed and solid 
lines are the model predictions without and with the absorption in the Milky
Way ISRF.}}
\label{fig:gamma-spect}
\end{figure*}

The DGE is produced through three major processes: decay of $\pi^0$
produced in $pp$-collisions, ICS and bremsstrahlung of CREs. At high
energies, the $\pi^0$ decay component dominates the DGE. Therefore we
only consider the $\pi^0$ decay component in the following calculation.
Comparisons between the model calculation and the measurements by ARGO-YBJ
\citep{2015ApJ...806...20B} and Tibet-AS$\gamma$ \citep{2021PhRvL.126n1101A}
are given in Fig.~\ref{fig:gamma-spect}, for two sky regions,
$25^\circ<l<100^\circ,~|b|<5^\circ$ and $50^\circ<l<200^\circ,~|b|<5^\circ$,
respectively. The DGE fluxes from the background sources are lower by a
factor of several than the data, as also shown in \citep{2021arXiv210403729Q}.
The inclusion of the secondary production from freshly accelerated CRs
interacting with the surrounding gas, which has a harder spectrum than
the CRs diffusing out, can reproduce the data well. 
{We can also estimate the interaction time of the source
component, which is about $5\times10^5$ years for the Galactic gas density 
distribution as adopted in DRAGON. It may be even shorter if some of the 
sources were located in denser molecular medium. This time reflects the 
confinement time of CRs in the vicinity of the sources.}
Note that at very high energies ($E\gtrsim100$ TeV), the absorption of $\gamma$ 
rays due to pair production with ISRF becomes important \citep{2006A&A...449..641Z},
which leads to a reduction of the DGE spectrum, as shown by the solid line.

\subsection{Ratios of B/C and $\bar p/p$}
\begin{figure*}[!htb]
\centering
\includegraphics[width=0.48\textwidth]{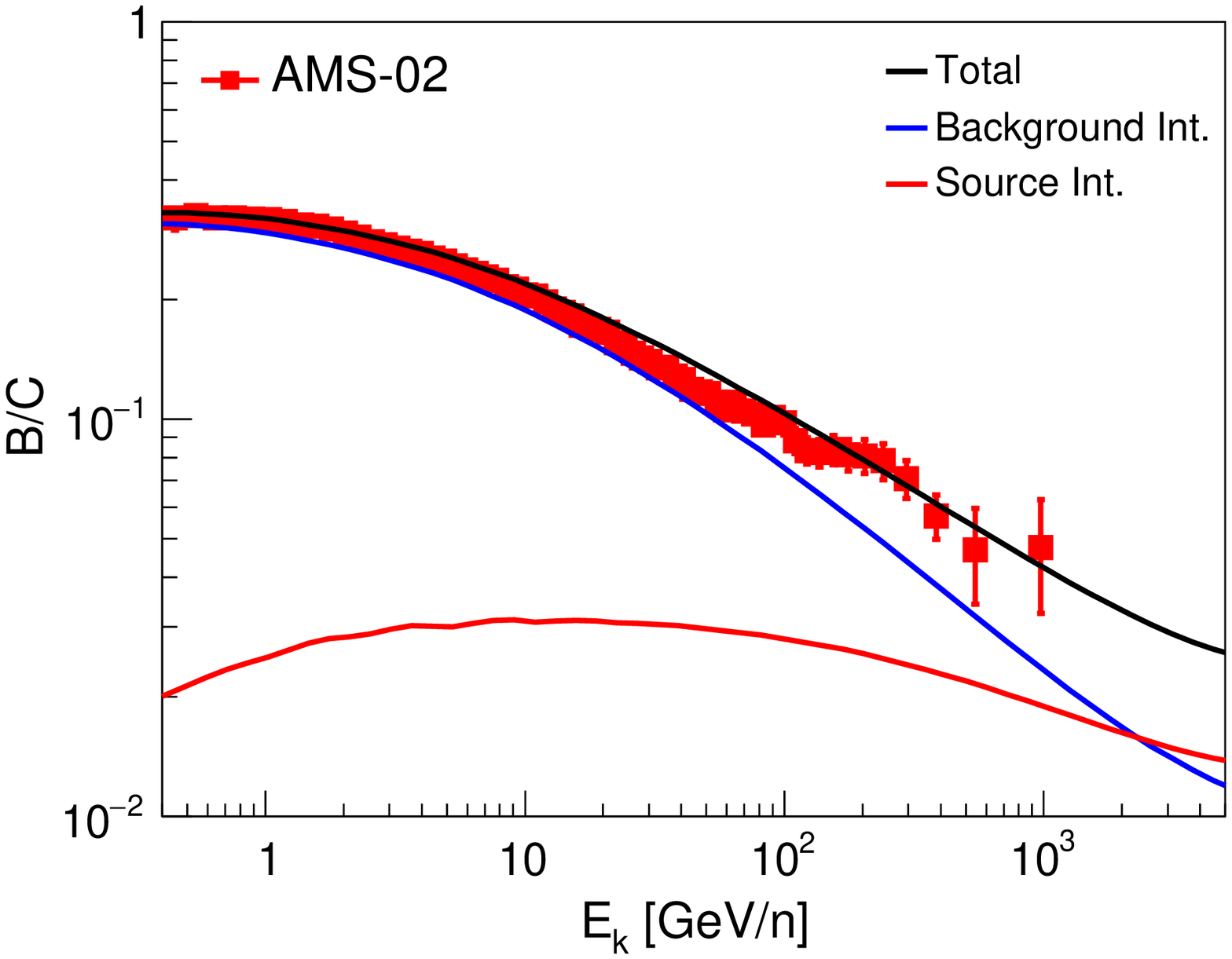}
\includegraphics[width=0.48\textwidth]{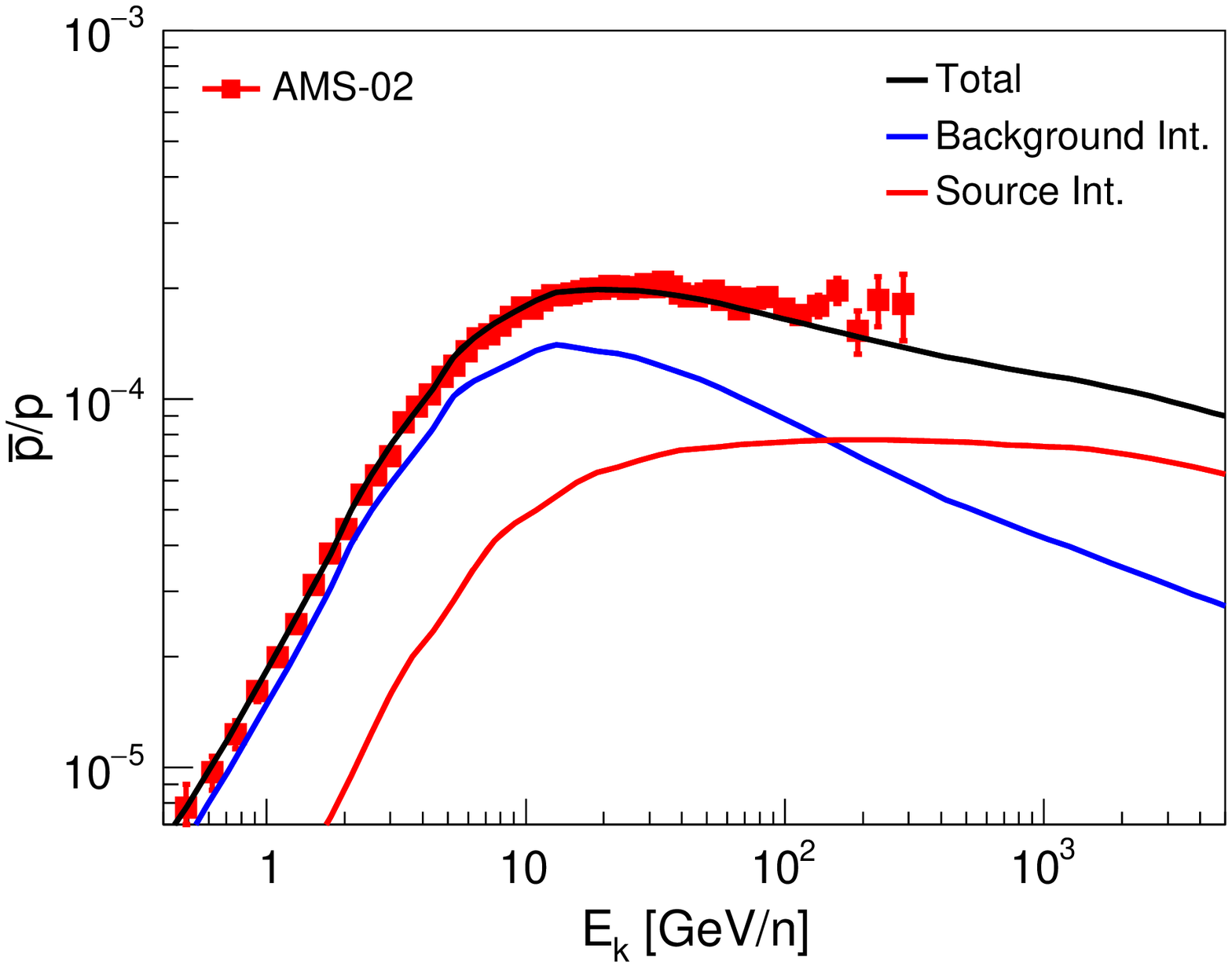}
\caption{The secondary-to-primary ratios of B/C (left) and $\bar p/p$ (right).
The data are from AMS-02 \citep{2016PhRvL.117w1102A,2021PhR...894....1A}.
}
\label{fig:BC-spect}
\end{figure*}

The same process to produce secondary $\gamma$ rays will generate
simultaneously secondary boron nuclei and antiprotons. The results of
the B/C and the $\bar p/p$ ratios are shown in Fig.~\ref{fig:BC-spect}.
Good consistency between the model and the data can be seen.
We note that the contribution of the ``fresh'' component exceeds the
background component when $E\gtrsim100$ GeV for $\gamma$ rays and
antiprotons, but it happens at much higher energies for B/C.
This is due to the fact that energies of secondary particles from
inelastic $pp$ interactions are much lower than those of parent
protons. However, for the nuclear fragmentation the kinetic energy
per nucleon keeps almost unchanged. 
{Note that for kinetic energies higher than $\sim100$ GeV, 
the measured $\bar{p}/p$ is slightly higher than the model prediction. 
Further refinement of the model parameters or additional source of 
antiprotons such as the dark matter annihilation 
\citep{2017PhRvD..95f3021H,2017PhRvD..96l3010L} may improve fitting to 
the data. This may also not be an issue due to the relatively large 
uncertainties of the measurements.
}


\subsection{Spectra of electrons and positrons}
\begin{figure*}[!htb]
\centering
\includegraphics[width=0.96\textwidth]{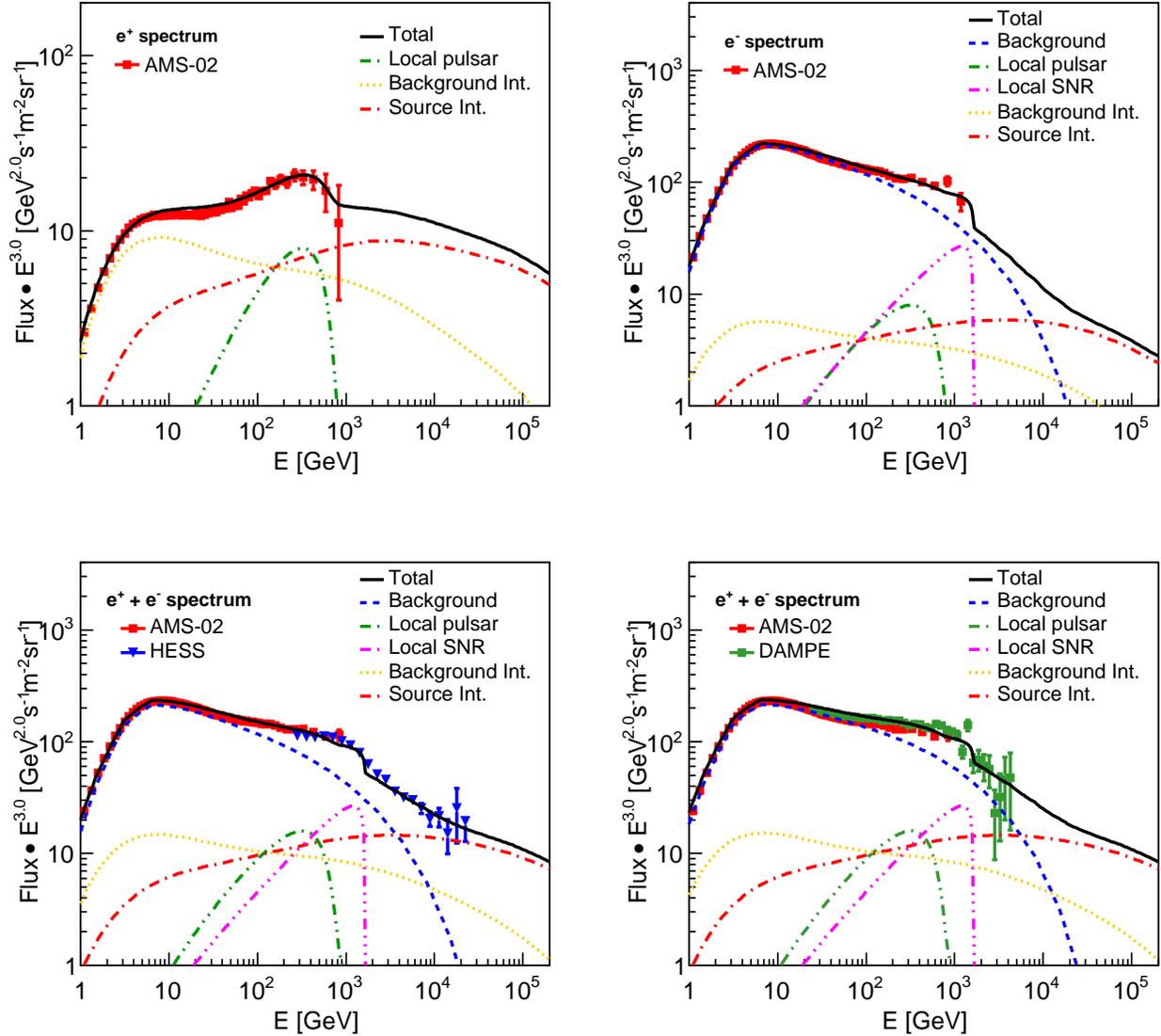}
\caption{The spectra of positrons and electrons (top) and total spectra
of positrons plus electrons (bottom). The measurements are from AMS-02
\citep{2019PhRvL.122j1101A,2019PhRvL.122d1102A}, H.E.S.S.
\citep{hess-icrc2017}, and DAMPE \citep{2017Natur.552...63D}.}
\label{fig:lepton-spect}
\end{figure*}

Finally we discuss the results of positrons and electrons. There are three
components of CR positrons, the secondary contributions from CRs interacting
when propagating in the Milky Way and around the acceleration sources,
and the primary contribution from the local pulsar. For CR electrons,
besides the same components as positrons, there are additional primary
components from both the background sources and the local SNR.
The results are given in Fig.~\ref{fig:lepton-spect}. Model parameters
of electrons are also given in Table \ref{table-parm}. For the total
CRE spectra, we give two groups of parameters according to the fittings
to the H.E.S.S. \citep{hess-icrc2017} and DAMPE \citep{2017Natur.552...63D}
data, which differ slightly. A clear feature of the model prediction is
that for energies above TeV, the fresh CR
interactions dominate the positron and electron spectra, resulting in
hardenings of their spectra. Such a property may be tested by further
precise measurements of the positron and electron spectra.

\begin{table*}
\begin{tabular}{|c|c|c|c|c|c|c|}
\toprule[1.5pt]
\hline
background  &  ${Q}_0~ [\rm m^{-2}sr^{-1}s^{-1}GeV^{-1}]^\dagger$  & $\nu_{1}$  & ${\cal R}_{\rm br}$ ~[GV] & $\nu_{2}$ & ${\cal R}_{\rm c}$ ~[GV]\\
\hline
$e^-$ for HESS fitting    & $2.70\times 10^{-1}$ & $1.14$ & $4.5$ & $2.77$  & $1\times 10^{4}$ &\\
$e^-$ for  DAMPE fitting   & $2.60\times 10^{-1}$ & $1.14$ & $4.5$ & $2.70$  & $1.1\times 10^{4}$ &\\
P    & $4.49\times 10^{-2}$ & $2.20$ & $7.2$ & $2.38$   &   $7\times 10^{6}$ &\\
He    & $3.74\times 10^{-3}$ & $2.20$ & $7.2$ & $2.32$  &   $7\times 10^{6}$ &\\
C    & $1.14\times 10^{-4}$ & $2.20$ & $7.2$ & $2.32$  &   $7\times 10^{6}$ &\\
N    & $9.18\times 10^{-6}$ & $2.20$ & $7.2$ & $2.35$  &   $7\times 10^{6}$ &\\
O    & $1.27\times 10^{-4}$ & $2.20$ & $7.2$ & $2.37$  &   $7\times 10^{6}$ &\\
Ne    & $1.62\times 10^{-5}$ & $2.20$ & $7.2$ & $2.30$ &   $7\times 10^{6}$ &\\
Mg   & $3.45\times 10^{-6}$ & $2.20$ & $7.2$ & $2.36$  &   $7\times 10^{6}$ &\\
Si   & $1.91\times 10^{-5}$ & $2.20$ & $7.2$ & $2.39$  &   $7\times 10^{6}$ &\\
Fe   & $1.82\times 10^{-5}$ & $2.20$ & $7.2$ & $2.31$  &   $7\times 10^{6}$ &\\
\hline

Local Pulsar &  $r_{\rm psr}$ [kpc] & $t_{\rm inj}$ [yrs] & $q_0^{\rm psr}~ [\rm GeV^{-1}]$  & $\gamma$  &  ${\cal R}_{\rm c}^{\rm e_\pm}$ ~[GV]  & $\tau_{0}$ [yrs]    \\

 & $0.33$  & $3.3\times 10^5$ & $2.7\times 10^{49}$  & $1.90$  & $800$  & $10^4$ \\

\hline
Local SNR &  $r_{\rm snr}$ [kpc] & $t_{\rm inj}$ [yrs] & $q_0^{\rm snr}~ [\rm GeV^{-1}]$  & $\gamma$ &  ${\cal R}_{\rm c}$ ~[GV]   &\\
\hline
$e^-$  & $0.33$ & $3.3 \times 10^5$ & $5.0\times 10^{49}$ &  $2.10$  & $2.8\times 10^{4}$ &    \\

P  & $0.33$ & $3.3 \times 10^5$ & $2.4\times 10^{52}$ &  $2.10$  & $2.8\times 10^{4}$ &    \\
He  & $0.33$ & $3.3 \times 10^5$ & $1.5\times 10^{52}$ &  $2.08$  & $2.8\times 10^{4}$ &    \\
C  & $0.33$ & $3.3 \times 10^5$ & $7.2\times 10^{50}$ &  $2.13$  & $2.8\times 10^{4}$ &    \\
N  & $0.33$ & $3.3 \times 10^5$ & $1.1\times 10^{50}$ &  $2.13$  & $2.8\times 10^{4}$ &    \\
O  & $0.33$ & $3.3 \times 10^5$ & $7.5\times 10^{50}$ &  $2.13$  & $2.8\times 10^{4}$ &    \\
Ne  & $0.33$ & $3.3 \times 10^5$ & $1.1\times 10^{50}$ &  $2.13$  & $2.8\times 10^{4}$ &   \\
Mg  & $0.33$ & $3.3 \times 10^5$ & $1.0\times 10^{50}$ &  $2.13$  & $2.8\times 10^{4}$ &   \\
Si  & $0.33$ & $3.3 \times 10^5$ & $1.0\times 10^{50}$ &  $2.13$  & $2.8\times 10^{4}$ &   \\
Fe  & $0.33$ & $3.3 \times 10^5$ & $1.8\times 10^{50}$ &  $2.13$  & $2.8\times 10^{4}$ &   \\

\toprule[1.5pt]
\end{tabular}\\
\caption{Injection parameters of different source components.
}
\label{table-parm}
\end{table*}
%


\section{Conclusion} \label{sec:concl}

The DGE at ultra-high energies is believed to be produced through the
interaction of CRs with the ISM, and is thus a good tracer to study
the propagation of galactic CRs. The ever first measurements of DGE
above 100 TeV energies by Tibet-AS$\gamma$ recently shows a significant
excess compared with the conventional CR propagation and interaction
model prediction. We find that possible hadronic interactions of CRs
with ambient gas surrounding the acceleration sources can account for
the ultra-high energy DGE by Tibet-AS$\gamma$. The harder spectrum of
CRs in the vicinity of the sources can naturally explain the high-energy
part of the DGE, while keeps the low-energy part unaffected.
The secondary interactions around the sources generate simultaneously
positrons and electrons, antiprotons, and boron nuclei. With proper
model parameters, we find that all these CR measurements can be well
reproduced. This model predicts hardenings of the spectra of both
positrons and electrons above TeV energies, and can be tested with
future measurements.

\begin{acknowledgments}
This work is supported by the National Key Research and Development Program
of China (No. 2018YFA0404203), the National Natural Science Foundation of
China (No.11875264, No.11635011, No.U1831208, No.U1738205).
Q.Y. is supported by the Key Research Program of the Chinese Academy of Sciences 
(No. XDPB15) and the Program for Innovative Talents and Entrepreneur in Jiangsu.

\end{acknowledgments}

\bibliographystyle{update}
\bibliography{refs}

\end{document}